\documentclass{raa}

\usepackage{graphicx,times}             %for PS/EPS graphics inclusion, new

\begin{document}

   \title{Strong screening effects on resonant nuclear reaction $^{23}$Mg $(p,\gamma)$ $^{24}$Al in the surface of magnetars$^*$
%\,$^*$
\footnotetext{$^{*}supported~ by~ the~ National~ Natural~ Science~
Foundation~ of~ China~ under~ grants~ 11565020, ~the~ Natural~
Science\\~ Foundation~ of~ Hainan~ province~ under~ grant~ 114012.$
}}
%   \subtitle{I. Place Your Subtitle Here}

   \volnopage{Vol.0 (201x) No.0, 000--000}      %%preserved for Editor. DOn't remove!
   \setcounter{page}{1}          %%starting page, preserved for Editor. DOn't remove!

   \author{Jing-Jing Liu\inst{}
   }
%% Here is an example of three authors come from different institutes.
%% For single author or all the authors from an institute, use "\inst{}" only

   \institute{College of Electronic and
Communication Engineering, Hainan Tropical Ocean University, Sanya,
572022,
China; {\it liujingjing68@126.com}\\
%% Please give the E-mail address of the author, to whom future correspondence and
%% offprint requests will be sent.
   }

\date{Received~~2014 month day; accepted~~2014~~month day}

\abstract{Based on the theory of relativistic superstrong magnetic
fields(SMFs), by using the method of the Thomas-Fermi-Dirac
approximations, we investigate the problem of strong electron
screening(SES) in SMFs, and the influence of SES on the nuclear
reaction of $^{23}$Mg $(p, \gamma)$$^{24}$Al. Our calculations show
that the nuclear reaction will be markedly effected by the SES in
SMFs in the surface of magnetars. Our calculated screening rates can
increase two orders of  magnitude due to SES in SMFs.
  }

   \authorrunning{Jing-Jing. Liu }            %author_head in even pages
   \titlerunning{An new estimate of electron capture of nuclides $^{55}$Co and $^{56}$Ni,}  % title_head in odd pages

   \maketitle
%% The author head (on even pages) and the title head (on odd pages) will be
%% automatically extracted from \author{} and \title{}. Whenever the title is too long,
%% you will be asked to supply a shorter one by inserting either \authorrunning{} or
%% \titlerunning{} before \maketitle. Anyway, you can specify your own heads.
%%
%%
%% Note: In the following text body of your manuscript, please note several differences from
%%       other major journals:
%% (1) \subsection{Please Capitalize the First Letter of Each Notional Word in Subsection Title}
%% (2) Please Capitalize the First Letter of Each Notional Word in all tables' captions

%
%________________________________________________ sections below
%
\section{Introduction}
According to the stellar evolution theory, for sufficient high
temperature in the Ne-Na cycle, the timescale of the proton capture
reaction of $^{23}$Mg is shorter than that of the $\beta^{+}$-decay.
Therefore, some $^{23}$Mg will kindle and escape from the Ne-Na
cycle by proton capture. The $^{23}$Mg leaks from the Ne-Na cycle
into the Mg-Al cycle and results in the synthesis of a large amount
of heavy nuclei. Thus the reaction rate of $^{23}$Mg $(p,\gamma)$
$^{24}$Al in stellar environment is of great importance to
nucleosynthesis of heavy nuclei. Due to its significance in
astrophysical surroundings, the nuclear reaction rate of $^{23}$Mg
$(p, \gamma)$ $^{24}$Al has been extensively studied. For instance,
by considering the contribution of a single resonance energy state,
Wallace et al. (1981) firstly discussed the reaction rate of
$^{23}$Mg $(p, \gamma)$ $^{24}$Al. Based on the three resonances and
a contribution from the direct capture process, Iliadis et al.(2001)
investigated this nuclear reaction rates. Taking into account four
resonances and the structure of $^{24}$Al, Kubono et al.(1995)
reconsidered the rate. Other authors (e.g., Herndl et al 1998;
Vissel et al. 2007; Lotay et al 2008) also carried out estimations
for the rate based on some new experimental information on $^{24}$Al
excitation energies. However, these authors seem to have overlooked
one important influence of electron screening on nuclear reaction in
a SMF.

The strong electron screening (SES) has always been a challenging
problem of the stellar weak-interaction rates and thermonuclear
reaction rates in pre-supernova stellar evolution and
nucleosynthesis. Some works (e.g., Bahcall et al. 2002; Liu
2013a,b,c,d, 2014a,b; 2015) have been done on stellar
weak-interaction rates and thermonuclear reaction rates. In
high-density plasma circumstances, the SES has been widely
investigated by various screened Coulomb model, such as Salpeter's
model (Salpeter et al. 1954, 1969), Graboske's model (Graboske et
al. 1973), Dewitt's model (Dewitt et al. 1976). The related
discussions were provided by Liolios et al. (2000), Liolios et al.
(2001), Kravchuk et al. (2014) and Liu (2013). Very recently,
Spitaleri \& Bertulani (2015) also discussed the electron screening
and nuclear clustering puzzle. Their results show that the large
screening potential values is in fact due to clusterization effects
in nuclear reactions, especially in reaction involving light nuclei.
However, they neglected the effects of SES on thermonuclear reaction
rate in SMFs. How does the SES influence the pre-supernova
explosion, nucleosynthesis and thermonuclear reaction in a SMF?  It
is very interesting and challenging for us to understand the
physical mechanism of SES in dense stars, especially in magnetars.

Magnetars have been proposed to be peculiar neutron stars which
could power their X-ray radiation by superstrong magnetic fields as
higer as  $B \sim 10^{14}- 10^{15}$ G (e.g., Peng \& Tong. 2007; Gao
et al. 2011a, 2011b, 2012; Guo et al. 2015; Xu \& Huang 2015; Xiong
et al. 2016) Some extensive researches about the characteristics,
emission properties, and the latest observations of magnetars have
been done. These researches on thermal and magnetic evolution of
magnetars are very interesting and challenging tasks in astronomy
and astrophysical environment. For instance, Tong (2015)
investigated the Galactic center magnetar J1745$-$2900 and gave a
note on the puzzling spin-down behavior. Olausen \& Kaspi (2014)
presented a catalog of the 28 known magnetars and candidates. They
investigated in detail their observed thermal radiative properties,
and the quiescent X-ray emission. Szary et al. (2015) discussed some
characteristics of radio emission from Magnetars. Based on the
estimated ages of potentially associated supernova remnants (SNRs)
of magnetars, Gao et al. (2016) discuss the values of the mean
braking indices of eight magnetars with SNRs. If the measurements of
the SNR ages are reliable, Gao et al. (2016) may provide an
effective way to constrain the magnetars' braking indices.

Recently, Li et al. (2016) numerically simulated the electron
fraction and electron Fermi energy in the interior of a common
neutron star. The electron Fermi energy and nuclear reaction rates
inside a magnetar will be affected substantially by SMFs (e.g., Gao
et al. 2011c, d; 2013, 2015).  In an extremely strong magnetic field
$(B\gg B_{\rm{cr}}, B_{\rm{cr}}=\frac{m^2_{\rm{e}}
c^3}{e\hbar}=4.414\times10^3$ G is the quantum critical magnetic
field)), the Landau column becomes a very long and very narrow
cylinder along the magnetic field. How does the quantization of
Landau levels change truly by SMFs? It is a very interesting issue
for us to discuss. Gao et al. (2013, 2015) investigated in detail
the pressure of degenerate for the relativistic electrons, and
discussed the quantization of Landau levels of electrons, and the
equations of states (EoSs) due to the quantum electrodynamic(QED)
effects for different matter systems by introducing Dirac
$\delta$-function in superhigh magnetic fields. Their results showed
that the stronger the magnetic field strength, the higher the
electron pressure becomes, and magnetars could be more compact and
massive neutron stars due to the contribution of magnetic field
energy.

In this paper, based on the SES theory in SMFs (Fushiki et al.
1989), we will carry out an estimation on the influence on the
electron Fermi energy, the SES and electron energy change due to
SMFs, and discuss the influence on the thermonuclear reaction by SES
in the surface of magnetars. Our work differs from previous works
(e.g., Peng \& Tong 2007; Gao et al. 2013, 2015) about the
discussion of electron Fermi energy in SMFs. Their works are based
on Pauli exclusion principle and Dirac $\delta$-function in
superhigh magnetic fields to discuss the influence of SMFs on the
electron Fermi energy and electron pressure. Although they discussed
the magnetic effects in detail, they have seemed to lose sight of
the influence of SMFs on SES. Following the works of Fushiki et al.
(1989), we will reinvestigate the electron Fermi energy in SMFs, and
derived new results for SES theory and the screening rates for
nuclear reaction in SMFs, based on the Thomas-Fermi-Dirac
approximations. Secondly, our discussions also differs from that of
Spitaleri \& Bertulani (2015), which analyzed the influence of the
SES only in the case without SMFs. Finally, Potekhin, \& Chabrier,
(2013) also discussed the electron screening effect on stellar
thermonuclear Fusion. However, they have just studied the impact of
plasma correlation effects on nonresonant thermonuclear reactions in
the liquid envelopes of neutron stars, and neglected the influence
of SES on resonant nuclear rates in SMFs.

The article is organized as follows. In the next Section, we will
discuss the properties of the free electron gas including the
electron Fermi energy and electron pressure in SMFs. Some
information of expressions of the SES in an SMF will be given in
Section 3. In Section 4, we will investigate the resonant reaction
process and rates in the case with and without SES and SMFs. In
Section 5, we will provide our main results and some discussions.
Section 6 gives  brief concluding remarks.

\section{The properties of the free electron gas in an SMF}

Theoretical studies of matter in high magnetic fields have been
carried out using a variety of methods, among which the Thomas-Fermi
(TF) and Thomas-Fermi-Dirac (TFD) approximations are the most used
methods, due to particular simple yet adequate for many purposes.
The TF method is the oldest and simplest case for a density
functional theory. The total energy of a system of electrons and
nuclei is written as a function of electron density. The detailed
investigations about the methods of TF and TFD approximations can be
referenced in Fushiki et al. (1991, 1992).

The positive electron energy levels, including the contributions of
its spin but neglecting radiative corrections in SMFs, are given
by (Landau \& Lifshitiz 1997)
\begin{equation}
E_n=n\hbar \omega_c+\frac{p_z^2}{2m_e},
 \label{1}
\end{equation}
where $n=0,1,2,....$, and $\hbar \omega_c=eB\hbar/m_ec\doteq
11.5B_{12}$ keV is the electron cyclotron energy, $B_{12}$ is the
magnetic fields in units of $10^{12}$G, $p_{\rm{z}}$ is the electron
momentum along the $z$-direction, and $m_e$ is the electron mass.
The electron chemical potential $U_e$ is determined by the inverting
expression for the electron number density
\begin{equation}
n_e=(\frac{eB}{hc})\frac{2}{h}[p_F(0)+2\sum_{n=0}^\infty
p_F(n)H(U_e-n\hbar \omega_c)],
 \label{2}
\end{equation}
where $p_F(n)=[2m_e(U_e-n\hbar\omega_c)]^{1/2}$ is the maximum
momentum along the $z$-direction for the n-th Landau orbit, $H(x)$
is the Heaviside function, which is unity when $x$ is positive and
is zero otherwise. By integrating Eq.(2) with respect to $U_e$, and
employing the Gibbs-Duhem relation, the pressure of electrons is
written as
\begin{equation}
P=(\frac{eB}{hc})\frac{2}{h}[\frac{p^3_F(0)}{3m_e}+2\sum_{n=1}^\infty
\frac{p^3_F(n)}{3m_e}H(U_e-n\hbar \omega_c)].
 \label{3}
\end{equation}

By summing over $n$, and integrating Eq.(1), the total kinetic
energy density, including contributions from the Landau orbit motion
perpendicular to the  field, the motion along the field and the
couping of the electron spin to the field, is
\begin{equation}
E_{kin}=(\frac{eB}{hc})\frac{2}{h}\{\frac{p^3_F(0)}{6m_e}+2\sum_{n=1}^\infty
[\frac{p^3_F(n)}{6m_e}+n\hbar \omega_c p_F(n)]H(U_e-n\hbar
\omega_c)\}.
 \label{4}
\end{equation}

According to the TFD approximations, the electron energy density
will include the contribution of electron exchange energy, and is
given by (Danz \& Glasser 1971)
\begin{equation}
E_{ex}=\frac{e^2}{2}(\frac{eB}{hc})^{-1}n_e^2F(\frac{n_e}{n_\ast})=\frac{r_{cyc}}{2\pi
a_0}\hbar\omega_c n_\ast n^2F(n),
 \label{5}
\end{equation}
where $a_0=5.29\times10^9$ cm is Bohr radius,
$n_\ast=2/\pi^{1/2}(eB/hc)^{3/2}=4.24\times 10^{27}B^{3/2}_{12}
\rm{cm}^{-3}$, $B_{12}$ is the magnetic fields in units of
$10^{12}$G, and $r_{cyc}=(2\hbar c/eB)^{1/2}\doteq3.63\times
10^{10}$ cm is the electron cyclotron radius in the lowest Landau
level.

From the TFD approximations, when only a single Landau level is
occupied, the electron chemical potential, which includes the
contribution of electron exchange energy, is determined by
\begin{equation}
U_{e}=\frac{\partial E_{ex}}{\partial
n_e}=e^2(\frac{eB}{hc})^{-1}n_eI(\frac{n_e}{n_\ast})=\frac{r_{cyc}}{\pi
a_0}\hbar\omega_c n_\ast nI(n),
 \label{6}
\end{equation}
where the expression of function $F(n)$ can be referenced in Fushiki
et al. (1989).

According to Eq.(1), the electron interaction energy with the
magnetic field is proportional to the quantum number $n$, and cannot
exceed the electron chemical potential. Thus the maximum Landau
level number $n_{max}$ will be related to the highest value of
interaction energy, allowed between electrons and the external
magnetic field. When $E( n_{max}, p_z = 0) = U_e$ in Eq. (1), the
maximum Landau level number
 $n_{max}$ will be given by
\begin{equation}
n_{max}=\frac{U_e}{\hbar \omega_c}.
 \label{7}
\end{equation}

In the general case, when $0\leq n\leq n_{max}$, the electron
momentum is less than its Fermi momentum $p_{F}(e)$, which is
determined by
\begin{equation}
p_{F}(e)= U_{e}/c,
 \label{7}
\end{equation}
 when $n=0$ for a superhigh magnetic field (e.g., Gao et al. 2013, 2015).

\section{The SES in a SMF}

According to Fushiki et al (1989), the nuclear reaction rate in
high-density matter is affected because the clouds of electrons
around nuclei alter the interactions among nuclei. Due to the
electron clouds, the reaction rate is increased by a factor of
$e^{U_{sc}/k_{B}T}$, where $U_{sc}$ is a negative quantity, called
``the screening potential'', and $T$ is the temperature. The
electron Coulomb energy by an amount which in the Wigner-Seitz
approximation in a SMF is given by
\begin{equation}
U_{\rm{sc}}=E_{\rm{atm}}(z_{12})-E_{\rm{atm}}(z_1)-E_{\rm{atm}}(z_2),
\label{8}
\end{equation}
where $E_{\rm{atm}}(z)$ is the total energy of Wigner-Seitz cell,
and $z_{12}=z_{1}+z_{2}$. If the electron distribution is rigid, the
contribution to $E_{\rm{atm}}(z)$ from the bulk electron energy
cancel, the electron screening potential at high density can be
expressed as
\begin{eqnarray}
U_{\rm{sc}}&=&E_{\rm{latt}}(z_{12})-E_{\rm{latt}}(z_1)-E_{\rm{latt}}(z_2)\nonumber\\
&=&\frac{-0.9e^2}{r_{\rm{e}}}[z_{12}^{5/3}-z_{1}^{5/3}-z_{2}^{5/3}],
 \label{9}
\end{eqnarray}
where $E_{\rm{latt}}(z)$ is the electrostatic energy of Wigner-Seitz
cell, $E_{atm}(z_{j})=\frac{-0.9z_{j}^{5/3}e^2}{r_{\rm{e}}}$, and
$r_e$ is radius of the Wigner-Seitz cell for a single electron. Due
to the influence of the compressibility of electron gas, the change
in screening potential is written as
\begin{eqnarray}
 \delta U_{\rm{s}}&=&-\frac{54}{175}(\frac{e^2}{r_{\rm{e}}})\frac{1}{n_{\rm{e}}}\frac{\partial n_{\rm{e}}}{\partial U_{\rm{e}}}[(z_{12})^{7/3}-(z_{1})^{7/3}-(z_{2})^{7/3}]\nonumber \\
 &=&-\frac{54}{175}(\frac{e^2}{r_{\rm{e}}})\frac{1}{n_{\rm{e}}}D[(z_{12})^{7/3}-(z_{1})^{7/3}-(z_{2})^{7/3}],
\label{10}
\end{eqnarray}
where
\begin{equation}
 D=823.1481\frac{r_{\rm{e}}n_{\rm{e}}}{e^2}(\overline{\frac{A}{z}})^{4/3}\rho^{-4/3}B_{12}^2.
\label{11}
\end{equation}
The Thomas-Fermi screening wavenumber will be given by
\begin{equation}
 (K_{\rm{TF}})^2=1.0344\times 10^4 r_{\rm{e}}n_{\rm{e}}(\overline{\frac{A}{z}})^{4/3}\rho^{-4/3}B_{12}^2.
\label{12}
\end{equation}
According to Fushiki et al. (1989), the corresponding change of
screening potential in a SMF is
\begin{eqnarray}
 \delta U_{\rm{s}}&=&-0.254(\overline{\frac{A}{z}})^{4/3}\rho^{-4/3}B_{12}^2 [(z_{12})^{7/3}-(z_{1})^{7/3}-(z_{2})^{7/3}] \nonumber \\
 &=&-494.668(\overline{\frac{A}{z}})^{4/3}\rho^{-4/3}b^2[(z_{12})^{7/3}-(z_{1})^{7/3}-(z_{2})^{7/3}]\rm{MeV},
\label{13}
\end{eqnarray}
where $(\overline{\frac{A}{z}})$ is the average of $\frac{A}{z}$
ratio, corresponding to the mean molecular weigh per electron, and
 $b=B/B_{\rm{cr}}=0.02266 B_{12}$. Thus, the electron screening
potential(hereafter ESP) in SMFs of FGP model is given by
\begin{equation}
U_{\rm{s}}=U_{\rm{sc}}+\delta U_{\rm{s}}.
\label{14}
\end{equation}
\section{The resonant reaction process and rates}
 \subsection{The calculation of resonant reaction rates with and without SES}
The reaction rates are contributed from the resonant and
non-resonant reactions. In the case of a narrow resonance, the
resonant cross section $\sigma_r$ is approximated by a Breit-Wigner
expression (Fowler et al. 1967)
\begin{equation}
 \sigma_{\rm{r}}(E)=\frac{\pi\omega}{\kappa^2}\frac{\Lambda_{i}(E)\Lambda_{f}(E)}{(E-E_{\rm{r}}^2)+\frac{\Lambda^2_{\rm{tot}}(E)}{4}},
\label{15}
\end{equation}
where $\kappa$ is the wave number, the entrance and exit channel
partial widths are $\Lambda_{i}(E)$ and $\Lambda_f(E)$,
respectively, $\Lambda_{\rm{tot}}(E)$ is the total width, and
$\omega$ is the statistical factor, which is given by
\begin{equation}
 \omega=(1+\delta_{12})\frac{2J+1}{(2J_1+1)(2J_2+1)},
\label{16}
\end{equation}
where the spins of the interacting nuclei and resonance are $J_1$,
and $J_2$, respectively, and $\delta_{12}$ is the Kronecker symbol.

The partial widths depend on the energy, and can be expressed as
(Lane et al. 1958)
\begin{equation}
\Lambda_{i,f}=2\vartheta_{i,f}^2\psi_{l}(E,a)=\Lambda_{i,f}\frac{\psi_{l}(E,a)}{\psi_{l}(E_{f},a)}.
\label{17}
\end{equation}
The penetration factor $\psi_{l}$ is associated with $l$ and $a$,
which are the relative angular momentum and the channel radius,
respectively, $a$ is written as $a=1.4(A_1^{1/3}+A_2^{1/3})$ fm.
$\Lambda_{i,f}$ is the partial energy widths at the resonance
process. $E_{\rm{r}}$ and $\vartheta_{i,f}^2$ is the reduced widths
and given by
\begin{equation}
 \vartheta_{i,f}^2=0.01\vartheta_{\rm{w}}^2=\frac{0.03\hbar^2}{2Aa^2}.
\label{18}
\end{equation}

Based on the above, in the phases of explosive stellar burning, the
narrow resonance reaction rates without SES are determined by
(Schatz wt al. 1998, Herndl et al. 1998)
\begin{equation}
 \lambda_{\rm{r}}^0=N_{\rm{A}}\langle{\sigma v}\rangle_{\rm{r}}=1.54\times10^{11}(AT_9)^{-3/2}\sum_{i} \omega \gamma_{i}\exp(-11.605E_{r_{i}}/T_9)  \rm{cm^3 mol^{-1} s^{-1}},
\label{19}
\end{equation}
where $N_{\rm{A}}$ is Avogadro's constant, $A$ is the reduced mass
of the two collision partners, $E_{r_i}$ is the resonance energies
and $T_9$ is the temperature in unit of $10^9$ K. The $\omega
\gamma_{i}$ is the resonance strength in units of MeV, and
determined by
\begin{equation}
 \omega \gamma_{i}=(1+\delta_{12})\frac{2J+1}{(2J_1+1)(2J_2+1)}\frac{\Lambda_{i}\Lambda_{f}}{\Lambda_{\rm{total}}}.
\label{20}
\end{equation}

On the other hand, due to SES, the reaction rates of narrow
resonance are given by
\begin{eqnarray}
 \lambda_{\rm{r}}^s&=&F_{\rm{r}} N_{\rm{A}}\langle{\sigma v}\rangle_{\rm{{r'}}}=1.54\times10^{11}(AT_9)^{-3/2}\sum_{i} \omega \gamma_{i}\exp(-11.605 E^{'}_{r_i}/T_9) \nonumber\\
 &=&1.54\times10^{11}F_{\rm{r}} (AT_9)^{-3/2}\sum_{i} \omega \gamma_{i} \exp(-11.605 E_{r_i}/T_9) ~~\rm{cm^3 mol^{-1} s^{-1}},
\label{21}
\end{eqnarray}
where $F_{\rm{r}}$ is the screening enhancement factor (hereafter
SEF). The values of $E^{'}_{r_i}$ should be measured by experiments,
but it is too hard to provide sufficient data. In a general and
approximate analysis, we have
$E^{'}_{r_i}=E_{r_i}-U_0=E_{r_i}-U_{\rm{s}}$.

\subsection{The screening model of resonant reaction rates in the
case with SMFs}

It is widely known that nuclear reaction rates at low energies play
a key role in energy generation in stars and the stellar
nucleosynthesis. The bare reaction rates are modified in stars by
the screening effects of free and bound electrons. The knowledge of
the bare nuclear reaction rates at low energies is important not
only for the understanding of various astrophysical nuclear
problems, but also for assessing the effects of host material in low
energy nuclear fusion reactions in stellar matter.

As mentioned in Section 1, most of manfnetars possess superhigh
surface dipole magnetic fields, and the internal magnetic field may
be higher than their surface magnetic field (e.g., Peng \& Tong
2007). Since the Fermi energy of the electron gas may go up to $10$
MeV, the quantum effects of electron gas will be very obvious and
sensitive to SMFs. The electron phase space will be strongly
modified by SMFs.  The electron screening will play a key role in
this process. It can strongly effect on the electron transformation
and nuclear reaction rates. In this subsection, we will discuss the
screening potential in the strong screening limit. The dimensionless
parameter $(\Gamma)$, which determines whether or not correlations
between two species of nuclei $(z_1, z_2)$ are important, is given
by
\begin{equation}
\Gamma=\frac{z_1z_2e^2}{(z_1^{1/3}+z_2^{1/3})r_{\rm{e}}kT}.
\label{22}
\end{equation}

Under the condition of $\Gamma\gg1$, the nuclear reaction rates will
be influenced appreciably by SES. The screening enhancement factor
(hereafter SEF) for resonant reaction process in SMFs can be
expressed as
\begin{equation}
F_{\rm{r}}^{\rm{B}}=\exp(\frac{11.605U_{\rm{s}}}{T_9}).
\label{23}
\end{equation}
\section{The results and discussions}

According to electron screening model of Fushiki et al. (1989) in
SMFs, we have calculated the electron screening potential at
deferent temperatures from Eqs.(10, 14, 15), based on the
Thomas-Fermi-Dirac approximations. Figure 1 shows that ESP is a
function of  $B_{12}$. We found that the SMFs have a slight
influence on the ESP in the high-density surroundings (e.g., $\rho_7
\geq 1.3$), but the influence on ESP is very remarkable for
relatively low densities (e.g., $\rho_7=0.1, 0.3, 0.5$) in SMFs. Due
to the fact that the higher the density, the lager the electron
energy becomes, it will definitely blunt the impact of SMFs on ESP.
For example, the ESP increases greatly when $B_{12}< 10^3$, and will
reach the maximum value of $0.0188$ MeV when $B_{12}=580.7$ and
$\rho_7= 0.1$. However, the ESP decreases about two orders of
magnitude when $10^3<B_{12}<2\times 10^3$ and $\rho_7=0.1$.

The influence of SES in SMFs on nuclear reaction is mainly reflected
by the factor of SEF. According to Eqs.(22, 24) and some parameters
of Table 1, we have calculated and analyzed tha factor of SEF in
detail. Figure 2 presents that the SEF is a function of magnetic
field strength $B$ for different temperature-density surroundings.
We find that the influences of SES on SEF are very remarkable in
SMFs. The lower the temperature, the greater the influence on SEF
becomes. This is because that the electron kinetic energy is
relatively low at lower temperatures. With the increasing of
magnetic field strength $B$, the SEF decreases. On the contrary, the
SEF increases greatly with increasing $B$ at relatively high
densities (e.g., $\rho_7= 1.0$).

Table 2 shows some important information about the SEF at cetain
astronomical conditions. We find that the lower the temperature, the
greater influence on the SEF. With the increasing of temperature at
the same density, the maximums value of SEF decreases. The maximums
value of SEF will get to 3289 when $B_{12}=10^4, \rho_7=10$ and
$T_9=0.1$. It is due to the fact that the higher the temperature,
the larger the electron energy. According to Eqs.(20, 22), we can
see that the nuclear reaction rates will increase as temperature
increases.

%      plot two figures side-by-side with epsf.sty
%------------------------------------------------------------ Fig 1:
\begin{figure*}
\centering
\includegraphics[width=12cm,height=12cm]{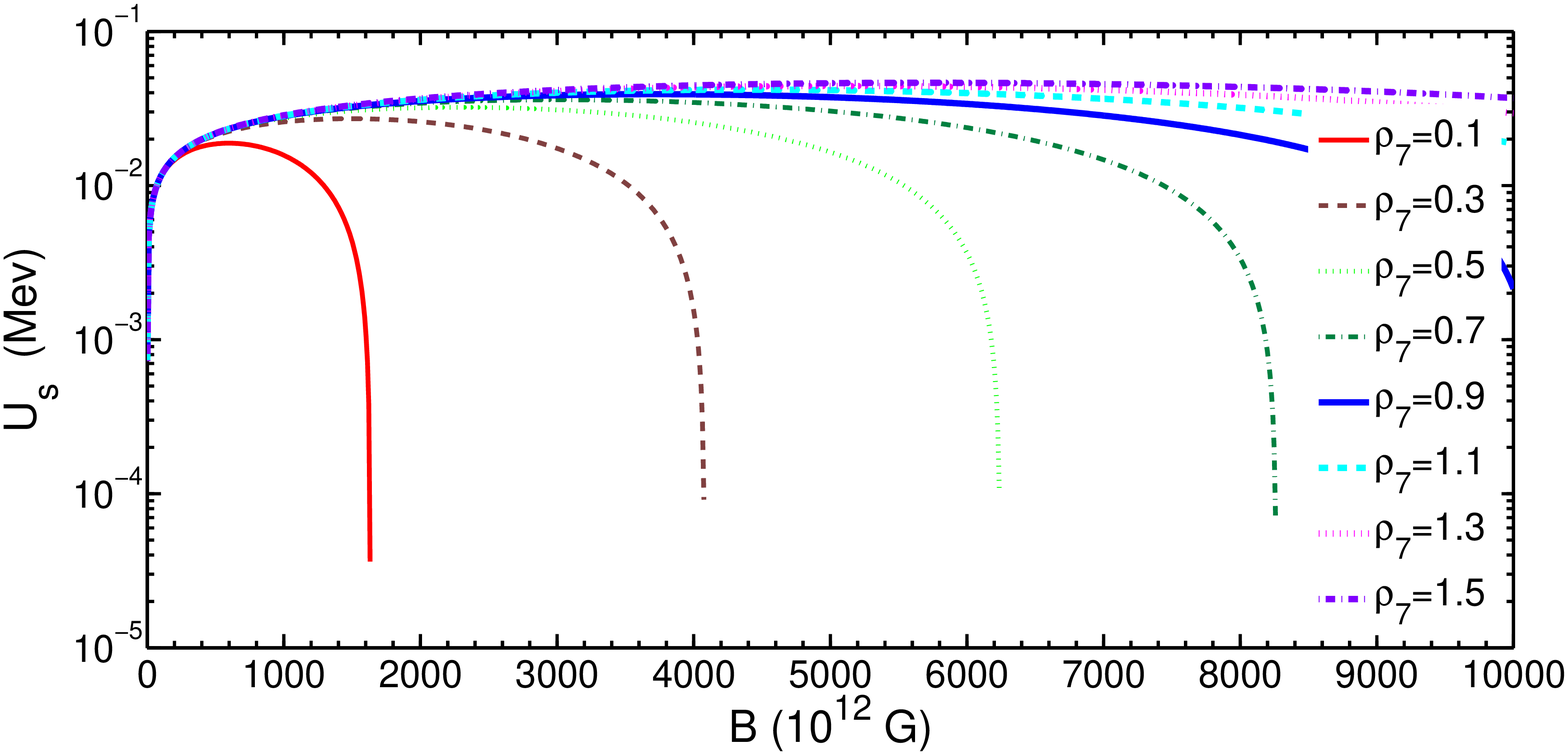}
\caption{The ESP as a function of $B$ under certain astronomical
conditions.} \label{fig1}
\end{figure*}
%      plot two figures side-by-side with epsf.sty
%------------------------------------------------------------ Fig 2:
\begin{figure*}
\centering
\includegraphics[width=7cm,height=7cm]{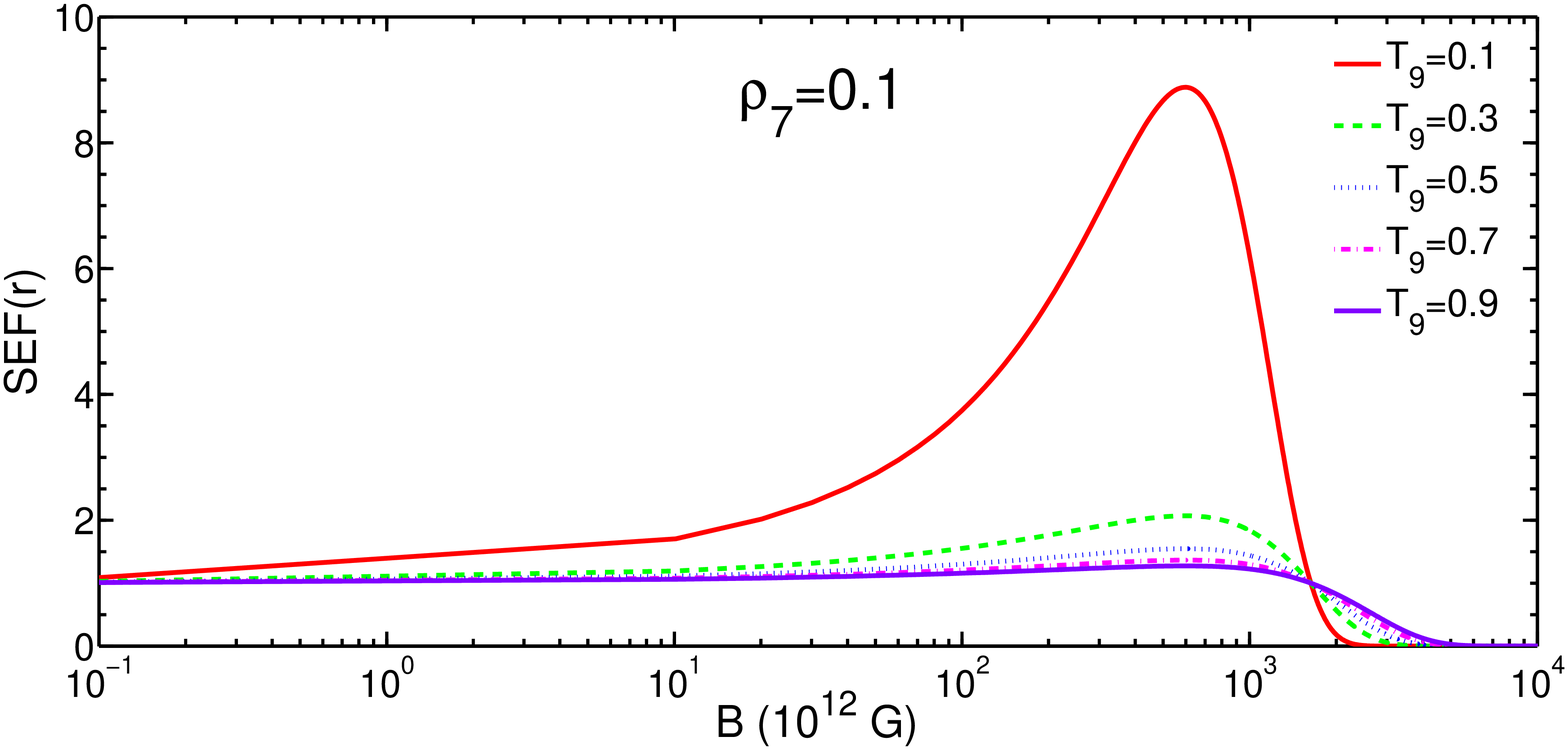}
\includegraphics[width=7cm,height=7cm]{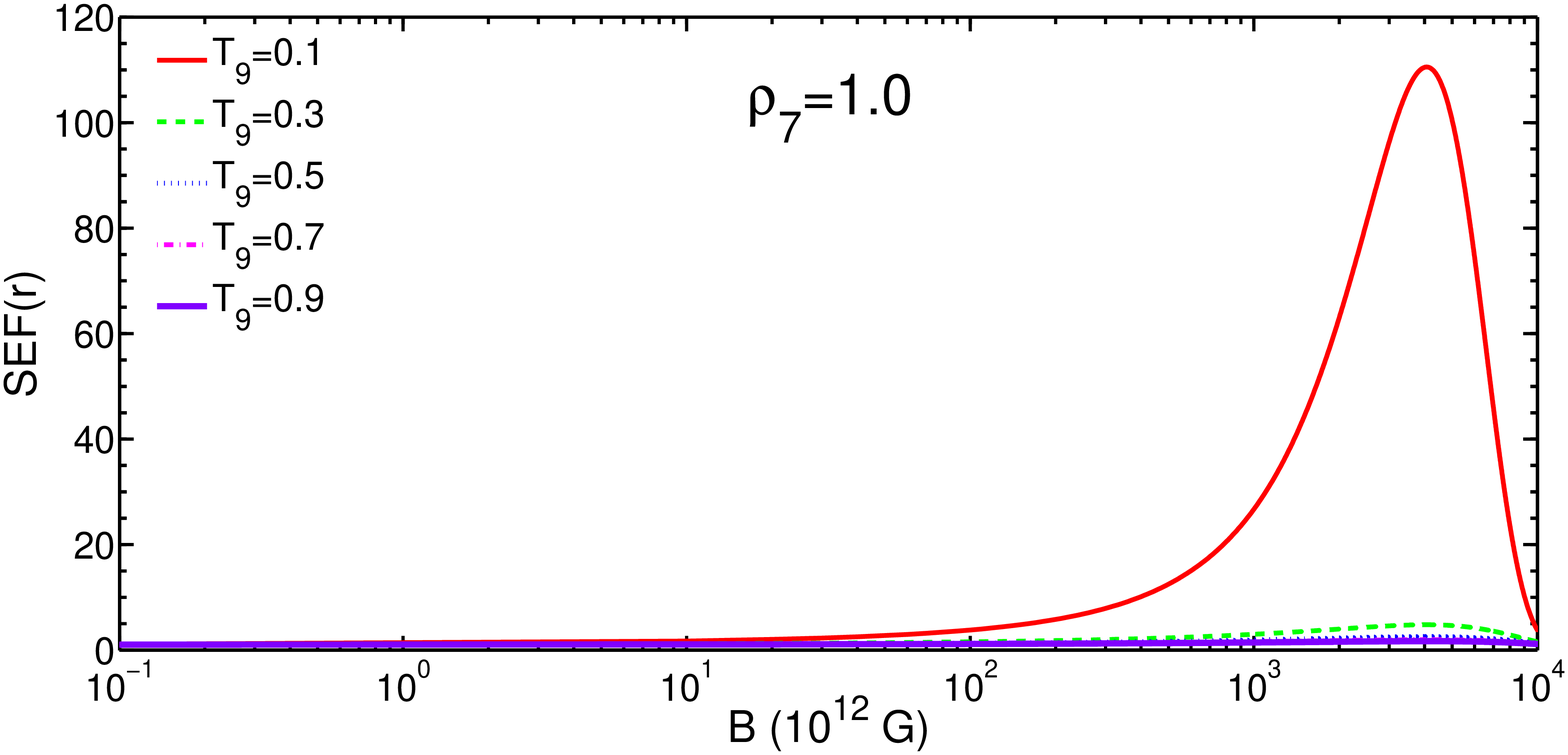}

\caption{The SEF as a function of $B$ for $\rho_7=0.1, 1.0$ under
certain astronomical conditions.} \label{fig2}
\end{figure*}

%      plot two figures side-by-side with epsf.sty
%------------------------------------------------------------ Fig 3:

\begin{figure*}
\centering
\includegraphics[width=7cm,height=7cm]{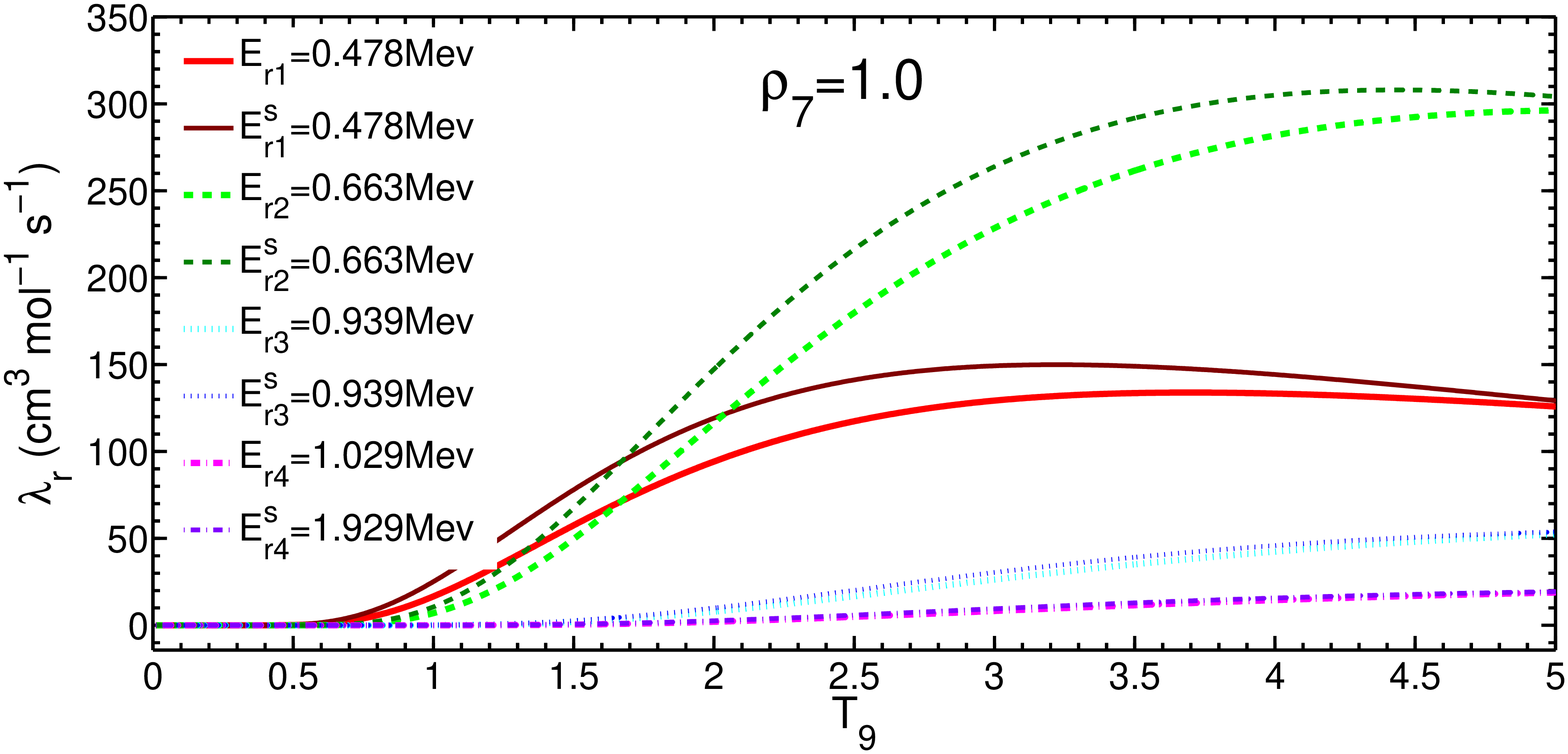}
\includegraphics[width=7cm,height=7cm]{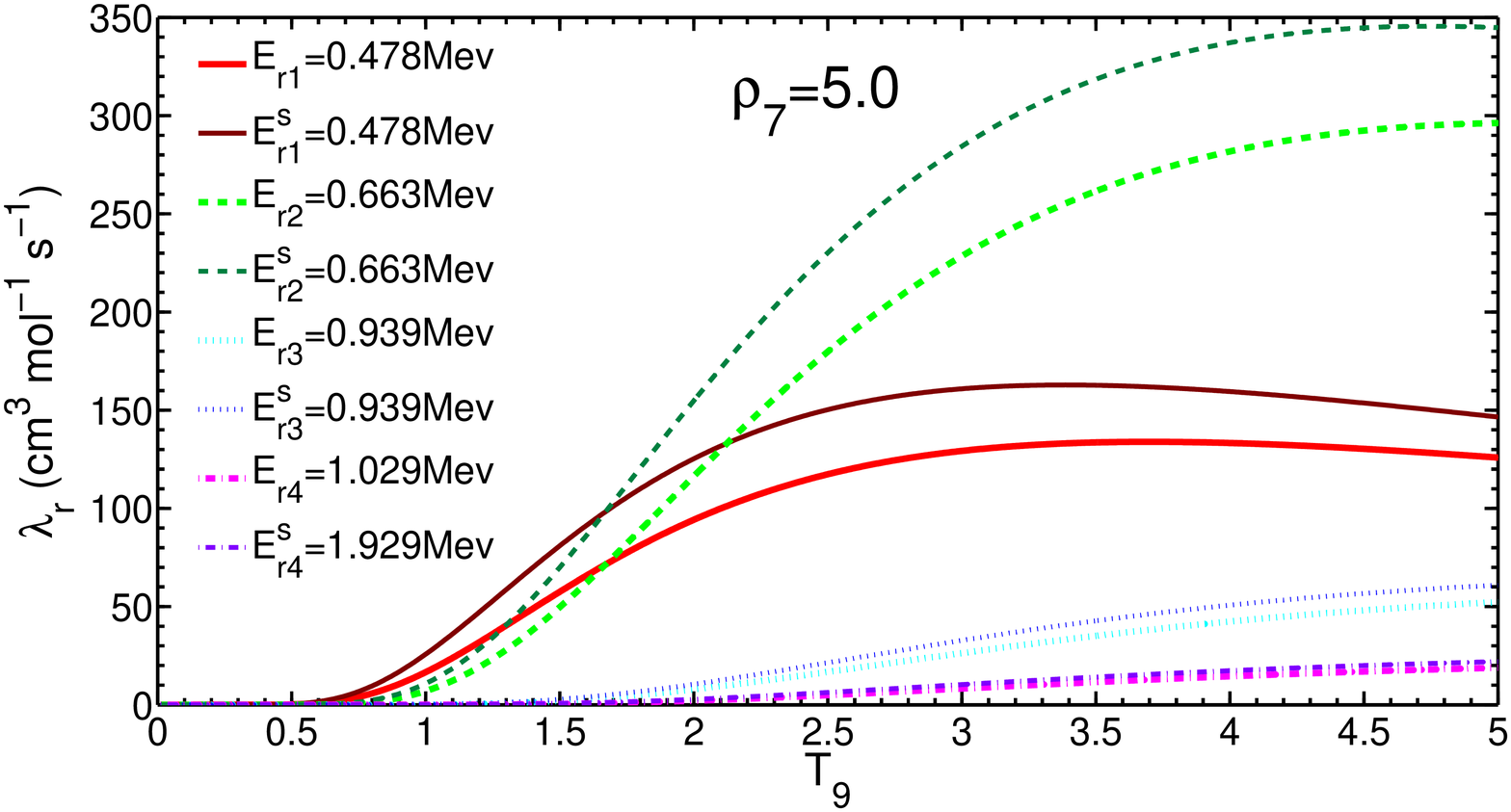}

\caption{The nuclear reaction rates in the case with and without SES
as a function of $T_{9}$ for $\rho_7=1.0, 5.0$ and $10^3 \rm{G} \leq
B \leq 10^{16}$ G in different energy states.} \label{fig3}
\end{figure*}
\clearpage
%________________________________________ Table 1:
\begin{table*}
\begin{minipage}{150mm}
\caption{Resonance parameters for the reaction $^{23}$Mg
$(p,\gamma)$ $^{24}$ Al.} \label{t.lbl}
\begin{center}
\begin{tabular}{@{}ccrrrrrrl@{}}
\hline

 $E_{\rm{x}}$ (MeV)  \footnote{is adopted from Ref. (Endt et al. 1998)} &$E_{\rm{x}}$ (MeV) \footnote{from Ref. (Visser et al.2007)}
 &$J^{\pi}$ & $E_{\rm{{r_{i}}}}$ (MeV) \footnote{ from Ref.(Audi et al. 1995)} & $\Gamma_{\rm{p}}$ &
 $\Gamma_{\gamma}$ & $\omega \gamma_{i}$(meV) \footnote{ from Ref.(Herndl et al. 1998)}
 &$\omega \gamma_{i}$(meV) \footnote{from Ref. (Wiescher et al. 1986)} & $\omega \gamma_{i}$(meV) \footnote{is adopted in this paper}\\
 \hline
2.349$\pm$0.020 &2.346$\pm$0.000 &$3^{+}$  &0.478   &185   &33  &25  &27  &26\\
2.534$\pm$0.013  &2.524$\pm$0.002 &$4^{+}$  &0.663   &2.5e3 &53  &58  &130 &94\\
2.810$\pm$0.020  &2.792$\pm$0.004 &$2^{+}$  &0.939   &9.5e5 &83  &52  &11  &31.5\\
2.900$\pm$0.020  &2.874$\pm$0.002 &$3^{+}$  &1.029   &3.4e4 &14  &12  &16  &14\\

\hline
\end{tabular}
\end{center}
\end{minipage}
\end{table*}

It is well known that, in explosive hydrogen burning stellar
environments, the nuclear reaction $^{23}$Mg$(p,\gamma)$$^{24}$Al
plays a key role in breaking out of the Ne-Na cycle to heavy nuclear
species (e.g., Mg-Al cycle). Therefore, it is very important for
accurately determinate the rates for the reaction $^{23}$Mg$(p,
\gamma)$$^{24}$Al. However, the resonance energy has a large
uncertainty due to inconsistent $^{24}$Mg($^3$He,t)$^{24}$Al
measurements, as mentioned above. Since different evaluation methods
may result in different sort orders, the evenness method is adopted
to increase accuracy of the comprehensive evaluation of Table 1.

According to Eqs.(22, 24) and some parameters of Table 1, the
resonant rates for four resonance states in the case with and
without SES are the functions of $T_{9}$, shown as in Figure 3. The
results show that the contributions of four resonant states to the
total reaction rate have obvious difference at the stellar
temperature range of $T_9 = 0.1- 5$. With the increasing of
temperature, the rates increase quickly. One can find that the
contribution of resonance state of $E_{\rm{r}}=478$ keV dominates
the total reaction rates when $T_9=0.2-1.681$, but the
$E_{\rm{r}}=663$ keV resonance is the most important at high
temperatures $T_9> 1.681$. On the contrary, the $E_{\rm{r}}=939$
keV, as well as $E_{\rm{r}}=1029$ keV resonance states are
negligible, compared to the former two lower resonance states over
the whole temperature range. Table 3 gives a brief description of
the resonant rates for four part resonance states due to SES in
SMFs. One can find that the maximum value of strongly screening
rates will reach 350.5 when $E_{2}=0.663$ MeV and $\rho_7= 100$.

In summary, by analyzing the influence of SES on resonant rates in
SMFs, we find that the SES has different effects on the rates for
different resonance states because of different forms of energy and
reaction orbits in the process of reaction in SMFs. We show that
this effect of SES is remarkable, and can increase reaction rates by
more than two orders of magnitude.

%________________________________________ Table 2:

\begin{table*}
 \caption{The maximums value of strong
screening enhance factor for some typical astronomical conditions.}
\centering
 \begin{minipage}{240mm}
  \begin{tabular}{@{}rrrrrrrrrrrr@{}}
  \hline
 & \multicolumn{2}{c}{$\rho_7=0.01$} & &\multicolumn{2}{c}{$\rho_7=0.1$}&&\multicolumn{2}{c}{$\rho_7=1.0$}&&\multicolumn{2}{c}{$\rho_7=10$}\\

\cline{2-3} \cline{5-6} \cline{8-9} \cline{11-12}\\
 $T_9$ &$B_{12}$ & $\rm{SEF}_{\rm{max}}$& & $B_{12}$ & $\rm{SEF}_{\rm{max}}$ & &$B_{12}$ & $\rm{SEF}_{\rm{max}}$& &$B_{12}$ & $\rm{SEF}_{\rm{max}}$  \\

 \hline
 0.1  &90.19 &2.755   & &590.7   &8.881   & &4074  &110.5     & &1000    &3289  \\
 0.3  &90.19 &1.402   & &610.7  &2.071   & &3954  &4.797     & &1000   &1487      \\
 0.5  &100.2 &1.223   & &650.7  &1.546   & &4074  &2.563     & &1000   &5.046   \\
 0.7  &110.2 &1.152   & &690.8  &1.362   & &4204  &1.985     & &1000   &3.179   \\
 0.9  &120.2 &1.113   & &630.7  &1.274   & &3914  &1.686     & &1000   &2.450  \\

\hline
\end{tabular}
\end{minipage}
\end{table*}

%________________________________________ Table 3:

\begin{table*}
 \caption{The \textbf{maximum} value of strong
screening enhance rates for some typical astronomical conditions.}
\centering
 \begin{minipage}{240mm}
  \begin{tabular}{@{}rrrrrrrrrrrr@{}}
  \hline
 & \multicolumn{2}{c}{$\rho_7=1.0$} & &\multicolumn{2}{c}{$\rho_7=5.0$}&&\multicolumn{2}{c}{$\rho_7=10$}&&\multicolumn{2}{c}{$\rho_7=100$}\\

\cline{2-3} \cline{5-6} \cline{8-9} \cline{11-12}\\
 $E$ &$\lambda^0_{\rm{max}}$ & $\lambda^s_{\rm{max}}$& & $\lambda^0_{\rm{max}}$ & $\lambda^s_{\rm{max}}$ & &$\lambda^0_{\rm{max}}$ & $\lambda^s_{\rm{max}}$& &$\lambda^0_{\rm{max}}$ & $\lambda^s_{\rm{max}}$  \\

 \hline
 $E_1=0.478$Mev  &131.9 &149.9   & &590.7   &8.881   & &133.7  &163.8     & &133.9    &164.7  \\
 $E_2=0.663$Mev  &296.2 &307.9   & &610.7  &2.071   & &296.2  &350.2     & &296.2   &350.5      \\
 $E_3=0.939$Mev  &51.50 &53.16   & &650.7  &1.546   & &52.31  &61.50   & &52.24   &61.90   \\
 $E_4=1.929$Mev  &18.45 &19.08   & &690.8  &1.362   & &18.59  &22.18     & &18.56   &22.27   \\

\hline
\end{tabular}
\end{minipage}
\end{table*}

\section{Concluding remarks}
The properties of matter in magnetars have always been interesting
and challenging objects for astronomers and physicists. The
investigation of SES is obviously an important component of magnetar
researches. In particular, improving the interpretation of nuclear
reaction data by SES in magnetars requires a detailed theoretical
understanding of physical properties for highly-magnetized nuclear
matter.

In this paper, employing the method of the Thomas-Fermi-Dirac
approximations in SMFs, we have investigated the problem of SES, and
the effects of SMFs on nuclear reaction of $^{23}$Mg $(p,
\gamma)$$^{24}$Al. Our calculations showed that the nuclear reaction
will be markedly affected by SES in SMFs of magnetars. The
calculated reaction rates can increase by more than two orders of
magnitude.  The considerable increase of reaction rates for
$^{23}$Mg $(p, \gamma)$ $^{24}$Al implies that more $^{23}$Mg will
escape the Ne-Na cycle due to SES, which will make the next reaction
convert more $^{24}$Al $(\beta^+, \nu)$ $^{24}$Mg to participate in
the Mg-Al cycle. It may lead to synthesizing a large amount of heavy
elements (e.g., $^{26}$Al) within the outer crust of magnetars.

\begin{acknowledgements}
We thank anonymous referee for carefully reading the manuscript and
providing valuable comments that improved this paper substantially.
This work is supported in part by Chinese National Science
Foundation through grant No.11565020, the Natural Science Foundation
of Hainan province under grant No.114012, and Undergraduate
Innovation Program of Hainan province under grant No.20130139.
\end{acknowledgements}

\label{lastpage}


\begin{thebibliography}{99}

\bibitem[1976]{Ashcroft76} Ashcroft, N. W., \&  Mermin, N. D., 1976, Solid State Physics (Saunders College: Philadelphia)
\bibitem[1995]{Audi95} Audi, G., \& Wapstra, A. H., 1995, Nucl. Phys. A., 595, 409
\bibitem[2002]{Bahcall02} Bahcall, J. N., Brown, L., Gruzinov, A., \& Sawer, R., 2002, A\&A, 383, 291
\bibitem[1968]{Canuto68} Canuto, V., \& Chiu, H. Y., 1968, Phys. Rev., 173, 1210
\bibitem[1971]{Canuto71} Canuto, V., \& Chiu, H. Y., 1971, Space. Sci. Rev., 12, 3
\bibitem[1976]{Dewitt76} Dewitt, H. E., 1976, Phys. Rev. A., 14, 1290
\bibitem[2012]{Das12} Das, Upasana., \& Mukhopadhyay, Banibrata., 2012, Phys. Rev. D., 86, 2001
\bibitem[1971]{Danz71}Danz, R. W.; Glasser, M. L.,1971, Phys. Rev. B., 4, 94
\bibitem[1998]{Endt98} Endt, P. M., 1998, Nucl. Phys. A., 633, 1
\bibitem[1967]{Fowler67} Fowler, W. A., Caughlan, G. R., \& Zimmerman, B. A., 1967, ARA\&A., 5, 525
\bibitem[1989]{Fushiki89} Fushiki, I., Gudmundsson, E. H., \& Pethick, C. J., 1989, \apj, 342, 958
\bibitem[1991]{Fushiki91} Fushiki, I., et al., 1991, Physics Letters A, 152, 1-2, p. 96
\bibitem[1992]{Fushiki92} Fushiki, I., et al., 1992, Annals of Physics, 216, 29
\bibitem[2011]{Gao11a} Gao, Z. F.; Peng, Q. H.; Wang, N.; et al., 2011a, Ap\&SS, 336, 427
\bibitem[2011]{Gao11b} Gao, Z. F.; Wang, N.; Song, D. L.; et al., 2011b, Ap\&SS, 334, 281
\bibitem[2011]{Gao11c} Gao, Z. F.; Wang, N.; Yuan, J. P.; Jet al., 2011c, Ap\&SS, 333, 427
\bibitem[2011]{Gao11d} Gao, Z. F.; Wang, N.; Yuan, J. P.; et al., 2011d, Ap\&SS, 332, 129
\bibitem[2012]{Gao12} Gao, Z. F.; Peng, Q. H.; Wang, N.; Yuan, J. P., 2012, Ap\&SS, 342,  55
\bibitem[2013]{Gao13} Gao, Z. F., et al., 2013, MPLA., 28, 1350138
\bibitem[2015]{Gao15} Gao, Z. F.; et al., 2015, Astron. Nachr. 336, 866
\bibitem[2016]{Gao16} Gao, Z. F., Li, X. D., Wang, N., et al., 2016, \mnras, 456, 55
\bibitem[1973]{Graboske73} Graboske, H. C., \& DeWitt, H. E., 1973, \apj, 181, 457
\bibitem[1983]{Gudmundsson83} Gudmundsson, E. H., et al., 1983, \apj, 272, 286
\bibitem[2015]{Guo15} Guo, Y-J., et al., 2015, RAA, 15, 525
\bibitem[1998]{Herndl98} Herndl, H., Fantini, M., Iliadis, C., Endt, P. M., \& Oberhummer, H., 1998, Phys. Rev. C., 58, 1798
\bibitem[2001]{Iliadis01} Iliadis, C., D'Auria, J. M., Starrfield, S., et al., 2001, \apjs, 134, 151
\bibitem[1970]{Kadomtsev70} Kadomtsev, B. B., 1970, Zh. Eksp. Teor. Fiz. 58, 1765
\bibitem[1965]{Kubo65} Kubo, R., 1965, Statistics Mechanics, (Amsterdam: North-Holland Publishing Co.), pp.~278, 280
\bibitem[1995]{Kubono95} Kubono, S., Kajino, T., \& Kato, S., 1995, Nucl. Phys. A., 588, 521
\bibitem[1977]{Landau77} Landau, L. D., \& Lifshitiz, E. M., 1977, Quantium mechanics, (3rd ed., Oxford: Pergamon Press), p.457
\bibitem[1958]{Lane58} Lane, A. M., \& Thomas, R. G., 1958, Rev. Mod. Phys., 73, 629
\bibitem[1985]{Lattimer85} Lattimer, J. M., et al., 1985, Nucl. Phys. A., 432, 646
\bibitem[2016]{Li16}Li X. H., et al., 2016, Int. J. Mod. Phys. D. 25(1), 1650002
\bibitem[2013]{Liu13a} Liu, J. J., 2013a, Research in Astronomy and Astrophysics, 13, 99
\bibitem[2013]{Liu13b} Liu, J. J., 2013b, Research in Astronomy and Astrophysics, 13, 945
\bibitem[2013]{Liu13c} Liu, J. J., 2013c, \mnras, 433, 1108
\bibitem[2014]{Liu14a} Liu, J. J., 2014a, Research in Astronomy and Astrophysics, 14, 971
\bibitem[2014]{Liu14b} Liu, J. J., 2014, \mnras, 438, 930
\bibitem[2014]{Liu15} Liu, J. J., 2015, Ap\&SS, 357, 93
\bibitem[2000]{Liolios00} Liolios, T. E., 2000, EPJA., 9, 287
\bibitem[2001]{Liolios01} Liolios, T. E., 2001, Phys. Rev. C., 64, 018801
\bibitem[2008]{Lotay08} Lotay, G., Wood, P. J., Seweryniak, D., et al., 2008, Phys. Rev. C., 77, 2802
\bibitem[2014]{Olausen14} Olausen, S. A.; Kaspi, V. M., 2014, \apjs, 212, 6
\bibitem[2003]{Pathria03} Pathria, R. K., 2003, Statistics Mechanics, 2nd., (Singapore: Isevier.), pp.~280
\bibitem[2007]{Peng07} Peng, Q. H., \& Tong, H., 2007, \mnras, 378, 159
\bibitem[2013]{Potekhin13} Potekhin, A. Y.; Chabrier, G.,2013, Contributions to Plasma Physics, 53, 397
\bibitem[1954]{Salpeter54} Salpeter, E. E., 1954, AuJPh., 7, 373
\bibitem[1969]{Salpeter69} Salpeter, E. E., \& van Horn, H. M., 1969, \apj, 155, 183
\bibitem[2015]{Spitaleri15}Spitaleri, C.; Bertulani, C. A., 2015, arXiv:1503.05266
\bibitem[1998]{Schatz98} Schatz, H., Aprahamian, A., Goerres, J., et al., 1998, Phys. Rep., 294, 167
\bibitem[1996]{Stolzmann96} Stolzmann, W., \& Bloecker, T., 1996, A\&A, 314, 1024
\bibitem[2015]{Szary15} Szary, Andrzej., et al.,2015, \apj, 800, 76
\bibitem[2015]{Tong15} Tong, H., 2015, RAA, 15, 1467
\bibitem[2007]{Visser07} Visser, D. W., Wrede, C., Caggiano, J. A., et al., 2007, Phys. Rev. C., 76, 065803
\bibitem[1981]{Wallace81} Wallace, R. K., \& Woosley, S. E., 1981, \apjs., 45, 389.
\bibitem[1986]{Wiescher86} Wiescher, M., et al., 1986, A\&A, 160, 56
\bibitem[2016]{Xiong16} Xiong, X-Y., Gao, C-Y, Xu, R-X., 2016, RAA, 16(1), DOI:10.1088/1674¨C4527/16/1/009
\bibitem[2015]{Xu15} Xu, M., \& Huang, Y-F., 2015, RAA, 15, 975
\bibitem[1989]{Yakovlev89} Yakovlev, D. G., \& Shalybkov, D. A., 1989, Astrophys. Space. Phys. Rev., 7, 311


\end{thebibliography}
\end{document}